\documentclass[preprint,12pt]{elsarticle}

\usepackage{amsmath}
\usepackage{amssymb}
\usepackage{graphicx}
\usepackage{bm}
\usepackage{subfig}
\usepackage{lineno}
\usepackage{microtype}
\usepackage[hidelinks]{hyperref}

\journal{Mechanical Systems and Signal Processing}

\begin{document}

\begin{frontmatter}

\title{A Reinforcement Learning Approach to Space-Filling Input Design for Nonlinear System Identification}

\author[vub]{Charly Witmeur\corref{cor1}}
\ead{charly.francois.r.witmeur@vub.be}
\author[vub]{Gonçalo G. Cruz}
\author[tue]{Maarten Schoukens}
\author[vub]{Mark~C.~Runacres}
\author[vub]{Jan Decuyper}

\cortext[cor1]{Corresponding author.}

\affiliation[vub]{organization={Vrije Universiteit Brussel (VUB)},
          addressline={Pleinlaan 2}, postcode={1050},
          city={Brussels}, country={Belgium}}
\affiliation[tue]{organization={Eindhoven University of Technology (TU/e)},
          addressline={PO Box 513}, postcode={5600 MB},
          city={Eindhoven}, country={The Netherlands}}

\begin{abstract}
In this paper, we propose a reinforcement learning (RL) approach to space-filling input design for nonlinear system identification, in which an agent learns a control policy that actively drives the system to explore its state space. Classical input design, developed mainly for linear models, minimises the covariance of the estimated parameters. For nonlinear black-box models, however, the dominant error is the bias induced by the model structure over the region of interest, which motivates space-filling strategies that seek to cover the system state space. Existing space-filling approaches rely on predefined input signal classes and require solving explicit optimisation problems. Unlike these methods, the proposed approach requires no input class to be selected and no frequency band to be tuned for the system at hand, satisfies an amplitude bound by construction, and adapts to varying initial conditions through its observation-based policy without re-optimisation. The learned excitation signals exhibit physically meaningful behaviours, such as resonance excitation and amplitude modulation. The approach is evaluated against tuned swept sine and crest-factor-optimised multisine excitations, and against a state-of-the-art space-filling multisine design, on two nonlinear benchmark systems: a mass-spring-damper with nonlinear stiffness and a Van der Pol oscillator. Under the same amplitude bound, the learned policy achieves the best overall space-filling performance on both systems and distributes its samples more uniformly than the state-of-the-art design.
\end{abstract}


\end{frontmatter}

\section{Introduction}\label{sec:intro}

Nonlinear system identification aims to capture a mathematical model of a dynamical system from its measured input and output data. While the identification of so-called linear time-invariant systems has achieved many successes over the past decades across a wide range of applications, real-world systems are inherently nonlinear and often time-varying, and ignoring these aspects is no longer appropriate in many contexts. Nonlinearities arise in mechanical engineering through nonlinear damping and stiffness, but also in biological~\cite{Korenberg1986}, telecommunication~\cite{Pedro2005} and fluid systems~\cite{DecuyperFluid}, to name but a few.

The recent boom in machine learning techniques has pushed modelling and parameter estimation forward, enabling the construction of increasingly flexible and powerful models. Successful identification, however, depends not only on the model structure and the parameter estimation method, but also on the data. Even the most flexible model cannot recover behaviour that the data does not contain, which increasingly makes the data the limiting factor for model quality. Improving the data means acting on the experiment itself, and in particular on the input applied to the system. For linear systems, the so-called optimal input design aims to minimise the variance of the identified model parameters induced by noise disturbances, typically through maximisation of the Fisher information matrix~\cite{mehra1974optimal}. However, for nonlinear systems, the paradigm shifts towards the bias, i.e., the model error arising from the chosen structure within a given model class and its limited ability to represent the nonlinear relationship between inputs and outputs over a given range~\cite{schoukens2019nonlinear}.

Indeed, black-box models are trained to reproduce the input-output relation of the training dataset, but correct identification of the system behaviour over the entire operational range is not guaranteed. While linear systems can be represented as a hyperplane in a high-dimensional space, nonlinear systems are characterised by a manifold in that same space. Intuitively, one can understand that correctly estimating this manifold requires covering it as uniformly as possible in the region of interest, rather than ensuring precise fitting of densely clustered points in a limited region. This requirement is not new: the restoring force surface method introduced by Masri and Caughey~\cite{masri1979nonparametric} reconstructs such a manifold explicitly. From a known mass and the measured excitation and acceleration, the restoring force is estimated as a surface over the system states. Because that surface is reconstructed on a regular grid while the excitation only produces data along a trajectory, the choice of excitation was quickly recognised as decisive: Worden~\cite{worden1990choice} compared several classes of excitation signal on that basis, judging them by how much of the phase plane the resulting trajectory covers. Designing the input so as to maximise that coverage is the idea behind space-filling input design. The space being covered depends on the model structure: it may correspond to the regressor space for NARMAX-type models, or to the joint input-state space for state-space models.

Designing a space-filling input is challenging for dynamical systems, as the state cannot be chosen directly but results from the initial condition and the whole past input sequence, which casts the problem as an optimisation over the input sequence or its parameters. Kiss et al.~\cite{KISS2024} successfully optimised the phases of a multisine signal, introducing a space-filling cost function within the optimisation procedure. Liu et al.~\cite{liu2025} extended this work by proposing a Gaussian process-based cost function. Other works~\cite{Nelles2017,Nelles2025,SMITS2025} have explored space-filling input design coupled with NARMAX model structures. These techniques, however, carry a number of limitations: they often rely on specific input classes or do not support directly enforce amplitude bounds, and require solving an explicit optimisation problem that must be repeated for each set of initial conditions. In this work, we propose the use of Reinforcement Learning (RL) as an alternative. Rather than optimising a fixed input signal, it learns a policy that maps system observations to inputs, which removes the need for a prescribed input parametrisation and for re-optimisation. While RL has already been applied to optimal experiment design, minimising parameter variance through Fisher-information-based criteria~\cite{Treloar2022,huang2023reinforcement}, it has not, to the best of our knowledge, been used to design inputs that fill the state space of nonlinear dynamical systems.

In this paper, we propose an approach to phase space-filling input design based on reinforcement learning that is general, independent of model structure and initial conditions. The main contributions are summarised as follows:
\begin{enumerate}
\item Space-filling input design is formulated as a reinforcement learning problem, in which a coverage reward defined on a discretisation of the region of interest in the state space is maximised over amplitude-bounded inputs, without prescribing an input parametrisation or tuning a frequency band for the specific system.
\item The learned policy is trained to generate inputs based on observations of the system. As a result, a single trained agent produces space-filling excitations from a range of initial conditions without re-optimisation.
\item The method is evaluated on two nonlinear benchmark systems against a swept sine, a crest-factor-optimised multisine and a state-of-the-art space-filling design~\cite{KISS2024}, and the learned inputs achieve competitive space-filling performance within the prescribed amplitude bound.
\end{enumerate}
The remainder of this paper is organised as follows. Section~\ref{sec:problem} formulates the space-filling input design problem. Section~\ref{sec:RL} details the proposed reinforcement learning framework. Section~\ref{sec:results} demonstrates the method on two nonlinear system simulations, and Section~\ref{sec:conclusion} concludes.

\section{Problem formulation}\label{sec:problem}
This paper develops an input design framework that tailors the excitation to the nonlinear system at hand, using a space-filling criterion. The input is designed offline, before any experiment on the real system, so that the design itself requires no costly interaction with it.

Evaluating such a criterion, however, requires the state trajectory induced by each candidate input, and hence a model of the very system that is to be identified, which makes the problem circular. In practice, this circularity is broken by a surrogate model, for instance identified from a preliminary experiment. In this paper, the surrogate is replaced by the exact system equations, in order to isolate the input design problem from surrogate model errors and to evaluate the framework purely from an input design perspective. Surrogate models are discussed in Section~\ref{sec:conclusion}.

We consider deterministic, time-invariant nonlinear systems in the discrete-time state-space form
\begin{align}
  \label{eq:state_space}
  \begin{aligned}
    \bm{x}(k+1) &= f(\bm{x}(k), \bm{u}(k)),\\
    \bm{y}(k)   &= g(\bm{x}(k), \bm{u}(k)),
  \end{aligned}
\end{align}
where $\bm{x}(k) \in \mathbb{R}^{n_x}$, $\bm{u}(k) \in \mathbb{R}^{n_u}$, and $\bm{y}(k) \in \mathbb{R}^{n_y}$ denote the state, input, and output vectors at time step $k \in \mathbb{N}_0$, a form that covers a wide range of data-driven models~\cite{Gulliver}. Since the design is carried out offline on a model, the state is fully available and noise-free, and no variance term enters the problem.

The functions $f$ and $g$ in Eq.~\eqref{eq:state_space} are evaluated at $\bm{z}(k) = [\bm{x}(k)^\top\ \bm{u}(k)^\top]^\top$, which lies in the joint input-state space. The objective is to design an input sequence $\{\bm{u}(k)\}_{k=0}^{N}$, with $\bm{u}(k) \in \mathcal{U} = [-u_{\max}, u_{\max}]^{n_u}$, such that, starting from an initial state $\bm{x}(0)$, the induced samples $\bm{z}(k)$ fill a user-defined region of interest $\mathcal{C}$ of that space as uniformly as possible. Since the induced trajectory depends on $\bm{x}(0)$, the design should ideally carry over to new initial states without re-optimisation.

For the two systems considered in Section~\ref{sec:results}, the input enters linearly through a constant matrix and the nonlinearity depends on the state only, so filling the state space suffices to excite the nonlinear behaviour. Therefore, in the remainder of this paper, $\bm{z}(k) = \bm{x}(k)$ and $\mathcal{C} \subset \mathbb{R}^{n_x}$. As no standard quantitative definition of this objective exists for dynamical systems, it is driven by a coverage reward during the design (Section~\ref{sec:reward}) and assessed by three complementary metrics (Section~\ref{sec:eval}).

For linear systems, good frequency and amplitude coverage of the input serves as an effective proxy for state-space coverage. For nonlinear systems, this proxy breaks down, and the induced state trajectory must be considered directly. The excitations commonly used in nonlinear system identification, such as swept sines and crest-factor-optimised multisines, have favourable frequency-amplitude properties. Whether they also induce good state-space coverage is, however, unclear. This question is addressed empirically in Section~\ref{sec:results}. Since these properties do not guarantee good coverage, there is no reason to restrict the input to a specific signal class. The intended operating conditions enter the problem only through the region of interest $\mathcal{C}$ and the amplitude bound $u_{\max}$, which typically reflects actuator limits, safety constraints or the intended operating range of the system. No other prior knowledge of input design is required. Section~\ref{sec:RL} describes how this problem is cast as a reinforcement learning task.

\section{Reinforcement learning input design}\label{sec:RL}
Although its objective is intuitive, space-filling input design is difficult to address with conventional optimisation methods, due to the high-dimensional continuous search space and the sequential dependence of the state trajectory on past inputs. Optimisation-based designs therefore typically restrict the input to a parametrised class, such as a multisine with optimised phases~\cite{KISS2024}, which keeps the problem tractable but limits the signals that can be designed.

\begin{figure}[t]
  \centering
  \includegraphics[width=\textwidth]{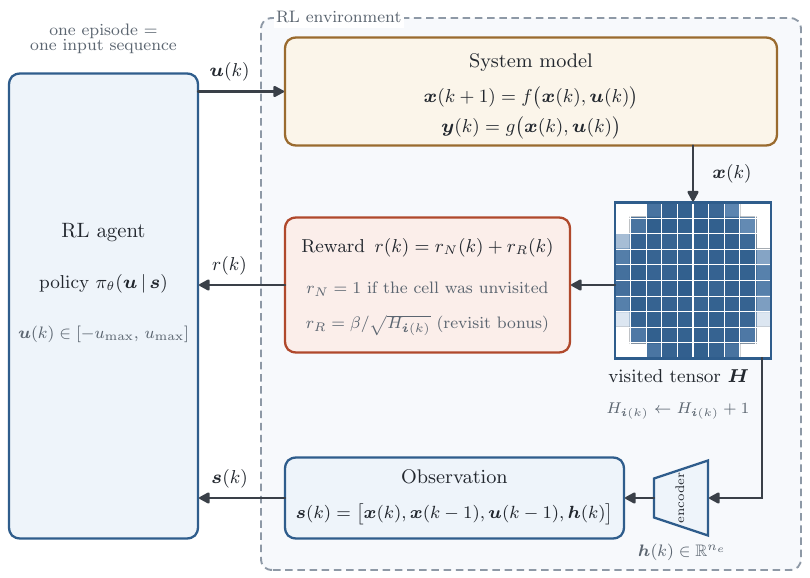}
  \caption{Overview of the proposed framework. At each time step, the agent selects an input sample within the amplitude bound. The system model is then advanced by one step, and the cell of the visited tensor $\bm{H}$ containing the resulting state is incremented. The reward combines a novelty term, granted when the cell was previously unvisited, with a count-based revisit bonus. The agent observes the current and past states, the past action, and a compressed representation of $\bm{H}$.}
  \label{fig:framework}
\end{figure}

Reinforcement learning is a branch of machine learning in which an agent learns to make decisions through trial and error, by interacting with an environment so as to maximise a cumulative reward~\cite{sutton2018reinforcement}. Because these decisions are made sequentially, each based on the current observation, RL matches the structure of the input design problem. An RL agent can thus learn a control policy that actively drives the system to generate space-filling trajectories, by interacting with a model of the system. The correspondence is direct, as summarised in Fig.~\ref{fig:framework}. The action $\bm{u}(k)$ selected by the agent at each step is the input applied to the system, and one episode corresponds to a complete input sequence of the prescribed experiment duration. The input is constructed sample by sample rather than parametrised in advance, and the reward scores the state trajectory that this sequence induces.

A successful RL agent depends on three components: the reward function guiding its behaviour, the observations provided to the agent, and the learning algorithm. These are specified in the following subsections.

\subsection{Reward function}\label{sec:reward}
In reinforcement learning, the agent is guided entirely by rewards rather than explicit solutions. The reward function therefore plays a dual role: it must reflect the space-filling objective, and it must deliver feedback in a way that guides the agent towards an optimal policy. In particular, sparse rewards (provided only at the end of an episode) are known to hinder learning, as the agent receives no signal to distinguish good actions from poor ones during the episode. Dense rewards, assigned at every time step, are therefore preferred. The reward at time step $k$ is accordingly defined as
\begin{equation}
r(k) = r_N(k) + r_R(k),
\end{equation}
where $r_N$ is a novelty reward that directly encodes the space-filling objective by incentivising the agent to visit previously unexplored regions, and $r_R$ is a revisit bonus, adapted from the count-based exploration bonus of~\cite{STREHL2008}, which keeps the reward informative when no new region is reached.

The novelty reward is computed as follows:
\begin{enumerate}
\item The region of interest $\mathcal{C}$ is discretised into a regular grid, stored in an $n_x$-dimensional tensor $\bm{H}$, called the visited tensor, initialised with zeros.
\item At each time step $k$, the current state $\bm{x}(k)$ is mapped to the index $\bm{i}(k)$ of the cell of $\bm{H}$ containing it.
\item The agent receives $r_N(k) = 1$ if $H_{\bm{i}(k)} = 0$, and $r_N(k) = 0$ otherwise.
\item The corresponding cell count $H_{\bm{i}(k)}$ is then incremented.
\end{enumerate}
The novelty reward accumulated over an episode therefore corresponds to the total number of unique cells visited, providing a direct measure of state-space coverage.

The revisit bonus is computed as
\begin{equation}
  r_R(k) = \frac{\beta}{\sqrt{H_{\bm{i}(k)}}},
\end{equation}
where $H_{\bm{i}(k)}$ is the visit count of the occupied cell after the update of step 4, so that $H_{\bm{i}(k)} \geq 1$. The scaling constant $\beta$ is set to $1$, so that the bonus of any revisit, at most $1/\sqrt{2}$, stays below the novelty reward of a new cell, preserving the primacy of the novelty reward while unexplored regions remain. This term encourages the agent to revisit less frequently explored regions, promoting uniformity of coverage.

\subsection{Observations}\label{sec:obs}
The choice of observations, that is the information available to the agent to perform its task, is equally fundamental. Too little information leads to sub-optimal agents, but too many observations can also be detrimental, making learning unnecessarily difficult.

As no standard observation set exists for space-filling input design, the observations were selected empirically, by comparing the training return reached with several candidate sets. The following observations were retained:
\begin{enumerate}
  \item Current state $\bm{x}(k)$,
  \item Past state $\bm{x}(k-1)$,
  \item Past action $\bm{u}(k-1)$,
  \item A compressed representation $\bm{h}(k) \in \mathbb{R}^{n_e}$ of the visited tensor $\bm{H}(k)$.
\end{enumerate}
They are concatenated into the observation vector
\begin{equation}
  \bm{s}(k) = \left[\bm{x}(k)^\top,\ \bm{x}(k-1)^\top,\ \bm{u}(k-1)^\top,\ \bm{h}(k)^\top\right]^\top,
\end{equation}
on which the policy $\pi_\theta(\bm{u}(k)\,|\,\bm{s}(k))$ is conditioned. The first three observations provide the agent with sufficient information to learn the system's state-space dynamics. The fourth conveys which regions of the state space have already been explored and which remain, allowing the agent, when paired with the revisit bonus of Section~\ref{sec:reward}, to prioritise poorly explored regions at each time step.

Although $\bm{H}(k)$ could be provided directly, its size, which grows exponentially with the number of states, would hinder learning. It is therefore compressed by the encoder of a multilayer perceptron autoencoder, trained offline prior to policy optimisation and kept fixed during training. The training set is built in two halves: tensors filled with random values within feasible bounds, which expose the autoencoder to a broad range of occupancy patterns, and tensors generated by applying classical excitation signals to the system, which are representative of the patterns encountered in practice. The latent dimension $n_e$ trades off preserving the information the agent needs against keeping the observation space small enough to learn efficiently; its value and the autoencoder architecture are given in Section~\ref{sec:implementation}.

\subsection{Learning algorithm: Proximal policy optimisation}
Proximal policy optimisation (PPO)~\cite{PPO} has emerged as one of the state-of-the-art reinforcement learning algorithms. In policy-based methods such as PPO, the agent aims to learn an optimal policy $\pi_\theta(\bm{u}(k) \mid \bm{s}(k))$, which defines the probability of taking an action $\bm{u}(k)$ given the current observation $\bm{s}(k)$ of the environment. In a deep reinforcement learning setting, the policy is represented by a neural network (the actor) parametrised by $\theta$. During training, these parameters are adjusted to maximise the expected discounted return over trajectories $\tau$:
\begin{equation}\label{eq:return}
  J(\theta) = \mathbb{E}_{\tau \sim \pi_\theta} \left[ R(\tau) \right],
  \qquad
  \text{with } R(\tau) = \sum_{k=0}^{N} \gamma^{k} r(k),
\end{equation}
where $\gamma$ is a discount factor weighting immediate versus future rewards.

Since actions are sampled from the policy during training, identical initial states may lead to different trajectories and returns, resulting in high variance in gradient estimates. Actor-critic methods, including PPO, mitigate this issue by introducing a value function $V(\bm{s}(k))$, estimated by a second neural network (the critic), which approximates the expected return from $\bm{s}(k)$. The advantage function, which quantifies how favourable an action is relative to this expected value, can be approximated using a one-step temporal-difference formulation:
\begin{equation}
  \hat{A}(k) = r(k) + \gamma V(\bm{s}(k+1)) - V(\bm{s}(k)).
\end{equation}
In the estimate of the gradient of $J(\theta)$, this advantage replaces the return $R(\tau)$, which reduces the variance of the estimate while leaving the objective unchanged.

PPO further improves training stability by introducing a clipped surrogate objective:
\begin{equation}\label{eq:clip}
  L^{\text{CLIP}}(\theta) = \mathbb{E}_k \Bigg[
     \min\!\Big(
        \rho_\theta(k)\, \hat{A}(k), \;
        \text{clip}\!\big(\rho_\theta(k), 1 - \epsilon, 1 + \epsilon\big)\,
        \hat{A}(k)
     \Big)
  \Bigg],
\end{equation}
where $\rho_\theta(k) = \pi_\theta(\bm{u}(k) \mid \bm{s}(k)) / \pi_{\theta_{\text{old}}}(\bm{u}(k) \mid \bm{s}(k))$ is the probability ratio of taking action $\bm{u}(k)$ given $\bm{s}(k)$ under the updated policy versus the policy at the previous update. The clipping removes any gradient incentive for updates that push $\rho_\theta(k)$ outside $[1-\epsilon,\,1+\epsilon]$, thereby preventing excessively large policy updates caused by inaccurate advantage estimates and stabilising training.

PPO has been widely adopted for continuous control problems and used as a reference algorithm in large-scale RL frameworks~\cite{openai-ppo}, due to its simple implementation, its efficiency, and the limited hyperparameter tuning it requires. Preliminary experiments with alternative RL algorithms were conducted, but PPO provided the most stable performance in our setting and is therefore retained.

\section{Results}\label{sec:results}
Two nonlinear systems are considered: a nonlinear mass-spring-damper system and the Van der Pol oscillator. The former exhibits nonlinear stiffness, while the latter is characterised by nonlinear damping, ensuring that the proposed approach is evaluated on two physically distinct classes of nonlinearity. Both are described by two-state nonlinear state-space models, enabling direct visualisation of state-space coverage while still exhibiting strong nonlinearities and distinct behaviours. They therefore provide a suitable benchmark for the proposed input design approach.

Both systems are simulated within the RL environment using discretised nonlinear state-space models of the form
\[
\bm{x}(k+1) = \bm{A}\bm{x}(k) + \bm{B}u(k) + \bm{E}\,\eta(\bm{x}(k)),
\]
where $\bm{x}(k) \in \mathbb{R}^2$ is the state vector, $\eta(\cdot)$ denotes system-specific nonlinear terms, and the sampling period is $T_s = 0.002\ \mathrm{s}$ for both systems.

\medskip

\textit{Mechanical system:} The nonlinear mass-spring-damper system is given by
\begin{equation}\label{eq:sys_mech}
\begin{aligned}
\bm{x}(k+1) &=
\begin{bmatrix}
1 & T_s \\
-\frac{s T_s}{m} & 1 - \frac{c T_s}{m}
\end{bmatrix}
\bm{x}(k)
+
\begin{bmatrix}
0 \\
\frac{T_s}{m}
\end{bmatrix}
u(k)
+
\begin{bmatrix}
0 \\
\frac{T_s}{m}
\end{bmatrix}
\phi(x_1(k)), \\
y(k) &= \begin{bmatrix}1 & 0\end{bmatrix}\bm{x}(k),
\end{aligned}
\end{equation}
where $\phi$ is the nonlinear part of the restoring force,
\[
\phi(x_1(k)) = \frac{s\, l\, x_1(k)}{\sqrt{x_1^2(k) + a^2}},
\]
with mass $m = 5\ \mathrm{kg}$, spring constant $s = 800\ \mathrm{N/m}$, damping coefficient $c = 10\ \mathrm{N\,s/m}$, tensionless length $l = 0.17\ \mathrm{m}$ and stretched length $a = 0.25\ \mathrm{m}$.

\medskip

\textit{Van der Pol system:} The forced Van der Pol oscillator is given by
\begin{equation}\label{eq:sys_vdp}
\begin{aligned}
\bm{x}(k+1) &=
\begin{bmatrix}
1 & T_s \\
-\omega_0^2 T_s & 1 + \varepsilon T_s
\end{bmatrix}
\bm{x}(k)
+
\begin{bmatrix}
0 \\
T_s
\end{bmatrix}
u(k)
+
\begin{bmatrix}
0 \\
-\varepsilon T_s
\end{bmatrix}
\psi(\bm{x}(k)), \\
y(k) &=
\begin{bmatrix}
1 & 0
\end{bmatrix}
\bm{x}(k),
\end{aligned}
\end{equation}
where the nonlinear term is $\psi(\bm{x}(k)) = x_1^2(k)\, x_2(k)$, with natural frequency $\omega_0 = \pi\ \mathrm{rad/s}$ and Van der Pol coefficient $\varepsilon = 1.5$.

\subsection{Implementation}\label{sec:implementation}
The PPO implementation from the Stable Baselines3 package (SB3)~\cite{sb3} is used. Default hyperparameters are retained, except for the discount factor $\gamma$ in  Eq.~\eqref{eq:return} and the clipping range $\epsilon$ in Eq.~\eqref{eq:clip}, set to $1$ and $0.1$ respectively. A discount factor of $1$ is chosen because episodes have a fixed, finite length and the objective is cumulative coverage: a cell visited late in the episode contributes as much as one visited early.

The environment follows the Gymnasium interface~\cite{gymnasium} and incorporates the system dynamics, reward function (Section~\ref{sec:reward}), and observations (Section~\ref{sec:obs}). The episode length is $20\,\mathrm{s}$, with a control update every $0.01\,\mathrm{s}$. System trajectories are recorded at each simulation time step, i.e. every $0.002\,\mathrm{s}$, resulting in $10\,000$ state samples per episode for the episode return and evaluation metrics described in Section~\ref{sec:eval}. The visited tensor is compressed to a latent dimension $n_e = 8$ by the autoencoder of Section~\ref{sec:obs}, a symmetric multilayer perceptron with two hidden layers of $128$ and $64$ units and ReLU activations on each side, and a sigmoid output layer.

The input amplitude lies in the continuous interval $[-400, 400]$ for both systems. The region of interest is defined as a rectangular domain, deliberately chosen larger than the expected trajectories to promote generalisation over the full admissible range. For the mechanical system, the region is $x_1 \in [-5,5]$, $x_2 \in [-50,50]$, and for the Van der Pol system $x_1 \in [-10,10]$, $x_2 \in [-70,70]$. In both cases, the region is discretised into a $30 \times 30$ grid, selected as a resolution that balances novelty-driven exploration and uniform state-space coverage. Training runs on $16$ parallelised environments for $10^{6}$ environment steps, approximately $15$ minutes on a standard laptop. The policy achieving the highest episode return during training is retained. PPO learns a stochastic policy; at evaluation the mean action is applied rather than a sample from it, so that the reported inputs are reproducible.

\subsection{Baseline excitation signals}\label{sec:baselines}
The RL inputs are compared with three baselines. The first two are standard excitation signals in nonlinear system identification, a swept sine and a \emph{crest-factor-optimised multisine}, both spanning $[0.1, 3]\,\mathrm{Hz}$. The upper bound of this range is independently selected for each system via a grid search over $[0.5, 10]\,\mathrm{Hz}$ to maximise state-space coverage, and happens to coincide at $3\,\mathrm{Hz}$, so that the agent is benchmarked against already optimised baselines. The crest-factor-optimised multisine has $59$ frequency lines of equal amplitude, with phases chosen to minimise the crest factor. Its standard deviation is set to $250$ so that its peak amplitude stays within the bound of $400$, like the swept sine and the RL input.

The third baseline, the \emph{space-filling multisine}, is the design of Kiss et al.~\cite{KISS2024}, the most closely related existing method. It optimises the phases of a multisine to minimise a kernel-density-based space-filling cost, evaluated here in the state space, the same space the proposed reward acts on, rather than in the joint input-state space of the original method. Its $59$ lines of equal amplitude cover the same frequency band, with the same standard deviation as the crest-factor-optimised multisine. Since the phases do not affect the standard deviation, the optimisation leaves the peak amplitude free. To enforce the amplitude bound of $400$ used throughout this paper, a hard constraint on the peak amplitude is therefore added to the space-filling cost, as the unconstrained designs, obtained with the RMS value of the RL input, reached peaks of $1533$ and $855$ for the two systems. All four inputs thus satisfy the same amplitude bound.

\subsection{Evaluation metrics}\label{sec:eval}
The concept of space-filling design for dynamical systems is still not fully standardised, and no universal metric exists. Nevertheless, two aspects are commonly used: minimising unfilled regions within the domain of interest, often characterised through the radius of the largest empty ball, and ensuring that the sampled points are distributed as uniformly as possible, for example through metrics such as the Jensen--Shannon divergence.

However, recalling that the objective of this work is to obtain a model that generalises well under any admissible input within a predefined maximum amplitude range, the extent of the covered region within the region of interest also plays a crucial role. To quantify this aspect, a domain coverage metric is introduced, defined as
\begin{equation}
C = \frac{A_{\mathrm{dom}}}{A_{\mathrm{ROI}}},
\end{equation}
where $A_{\mathrm{dom}}$ denotes the area of the convex hull of the sampled state-space points, and $A_{\mathrm{ROI}}$ is the area of the region of interest.

In addition, the sampled state-space points are used to construct a Voronoi diagram~\cite{Voronoi}, also known as a Dirichlet tessellation, which partitions the plane into convex cells such that each cell contains exactly one generating point and every point inside a given cell is closer to its generating point than to any other. The diagram is computed on the state samples normalised by the size of the region of interest along each state variable, so that metrics do not depend on the different scales of the states.

Using the statistics of these Voronoi cells, two complementary metrics are extracted. First, the worst uncovered region is quantified through the relative maximum Voronoi cell area
\begin{equation}
M = \frac{\max_i A_i}{A_{\mathrm{dom}}},
\end{equation}
where $A_i$ denotes the area of the $i$-th Voronoi cell. Second, the coefficient of variation (CV) of the Voronoi cell areas is used as a metric of coverage uniformity:
\begin{equation}
\mathrm{CV} = \frac{\sigma_A}{\mu_A},
\end{equation}
where $\sigma_A$ and $\mu_A$ denote the standard deviation and mean of the Voronoi cell areas, respectively.

\begin{figure}[htbp]
  \centering
  \includegraphics[width=\textwidth]{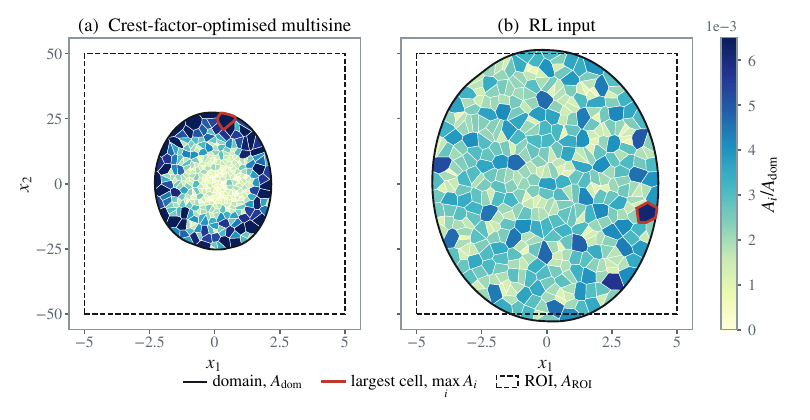}
  \caption{Voronoi tessellation of the sampled state-space points for two excitations on the mechanical system. The convex hull of the samples defines $A_{\mathrm{dom}}$, whose ratio to the area of the region of interest (dashed box) gives the domain coverage $C$. Each cell is shaded by $A_i/A_{\mathrm{dom}}$, so that uniform shading corresponds to a low coefficient of variation $\mathrm{CV}$ and the darkest cell, outlined in red, is the one defining $M$ for the tessellation shown. For legibility, the tessellation is drawn for every 25th sample. The metrics in Table~\ref{tab:perf_mech} are computed on the complete trajectories.}
  \label{fig:eval_metric}
\end{figure}

Since different excitation signals achieve different levels of state-space coverage, the absolute Voronoi cell areas are not directly comparable: for a fixed number of samples, a larger explored region yields larger cells. The maximum cell area is therefore normalised by the domain area $A_{\mathrm{dom}}$, so that $M$ measures the largest uncovered region relative to the region actually explored. Likewise, normalising the standard deviation of the cell areas by their mean $\mu_A$ removes the dependence on the overall cell size and yields the dimensionless coefficient of variation $\mathrm{CV}$, which quantifies relative non-uniformity. Lower values of $M$ and $\mathrm{CV}$ indicate better space-filling properties in terms of local coverage and uniformity, and the three metrics are complementary and must be read jointly. In particular, $M$ only reflects the largest uncovered region, whereas $\mathrm{CV}$ accounts for all cells. A high $M$ combined with a low $\mathrm{CV}$ therefore indicates one or a few large gaps in an otherwise evenly covered domain, whereas a similar $M$ with a higher $\mathrm{CV}$ indicates that several large gaps are present. Fig.~\ref{fig:eval_metric} illustrates them on two of the inputs applied to the mechanical system.

These metrics are used exclusively for evaluation: as they require the complete state trajectory to be computed, they cannot provide the dense per-step feedback necessary for effective RL training, for which the visited tensor-based reward described in Section~\ref{sec:reward} serves as a tractable proxy.

\subsection{Space-filling performance}
\subsubsection{Mechanical system}
\begin{figure}[p]
  \centering
  \includegraphics[width=\textwidth]{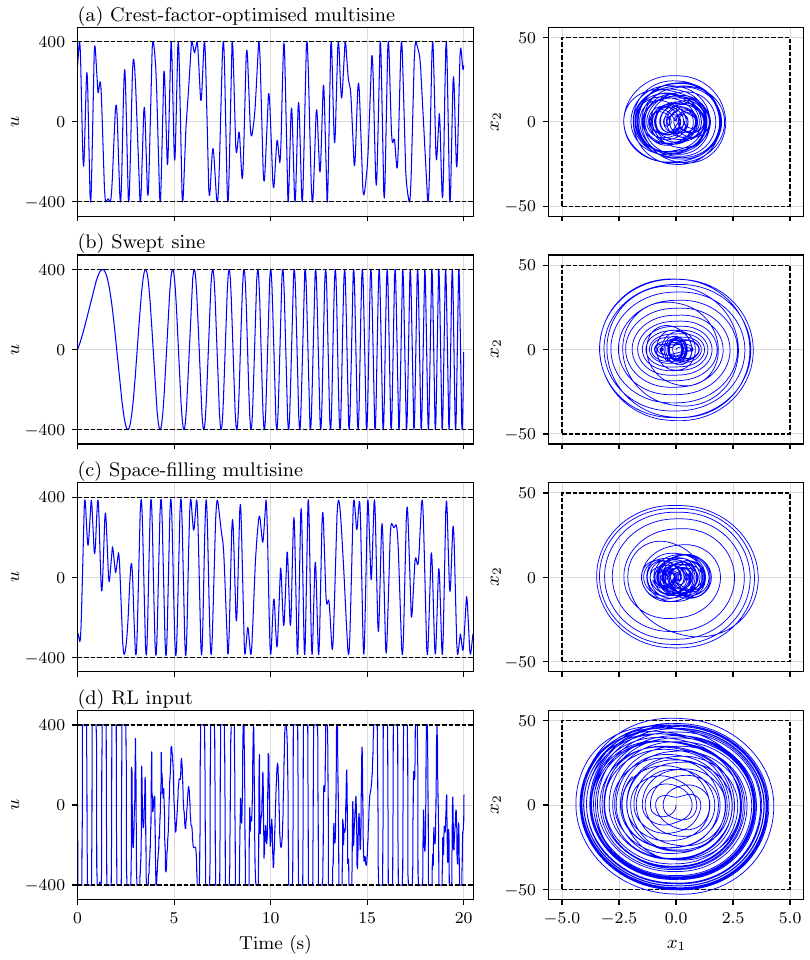}
  \caption{Input signals (left) and the state-space trajectories they induce (right) on the mechanical system. The dashed lines mark the amplitude bound and the region of interest.}
  \label{fig:mech_signals}
\end{figure}

Fig.~\ref{fig:mech_signals} shows the input signals (left) and the state-space trajectories they induce (right) for the crest-factor-optimised multisine (a), the swept sine (b), the space-filling multisine (c) and the RL input (d). Table~\ref{tab:perf_mech} reports the corresponding metrics.

The crest-factor-optimised multisine keeps the trajectory in a small region around the centre of the ROI, where its samples concentrate, giving the lowest domain coverage, $C = 18.79\%$. Optimising the phases for space filling enlarges this region considerably. The space-filling multisine reaches $C = 47.34\%$, slightly above the $44.97\%$ of the swept sine, but with many large gaps inside it, which raises the coefficient of variation from $1.09$ to $1.99$, the highest of the four. The swept sine leaves the smallest largest gap, $M = 0.67\times10^{-3}$, but many of its samples cluster near the centre, giving a coefficient of variation of $1.44$.

The RL input covers a much larger part of the state space under the same amplitude bound and distributes its samples far more evenly. It achieves the highest cumulative reward and the largest domain coverage, $C = 71.72\%$, together with the lowest coefficient of variation, $0.52$, less than half that of every baseline. Its worst uncovered ratio $M$ is nevertheless the highest of the four, because of a single gap at the centre of the ROI, which the trajectory leaves at the start and does not revisit (Fig.~\ref{fig:mech_signals}d). The rest of the domain is covered evenly, hence the low $\mathrm{CV}$. The space-filling multisine shows the opposite pattern: its largest gap is smaller, but it has many of them, in a sparsely covered annulus between its dense core and its outer orbits (Fig.~\ref{fig:mech_signals}c), hence its high $\mathrm{CV}$.

Interestingly, the learned input takes the form of an amplitude-modulated oscillatory signal. The agent appears to have autonomously identified the dominant resonance frequency of the nonlinear system under a $400\,\mathrm{N}$ excitation, at approximately $2\,\mathrm{Hz}$. At such amplitudes, $\phi$ saturates at $\pm s l$, so that the slope of the restoring force tends to $s$ and the resonance approaches $\sqrt{s/m}/(2\pi) \approx 2.0\,\mathrm{Hz}$. Furthermore, the amplitude modulation enables the agent to progressively explore both low- and high-energy regions of the state space, thereby improving coverage uniformity. This suggests that the RL agent is not merely maximising reward heuristically, but is learning physically meaningful excitation strategies. Notably, the amplitude-modulated form it converges to coincides with an excitation known to be effective for restoring force surface identification~\cite{worden1990choice}, which the agent recovers without any prior on the input structure.

\begin{table}[htbp]
\centering
\caption{State-space coverage performance on the mass-spring-damper system. A high return and domain coverage $C$ indicate a wide exploration of the region of interest, a low $M$ the absence of large uncovered regions, and a low $\mathrm{CV}$ a uniform distribution of samples.}
\label{tab:perf_mech}
\small
\begin{tabular}{lcccc}
\hline
\textbf{Input} & \textbf{Return}$\uparrow$ & $\bm{C}$ \textbf{(\%)}$\uparrow$ & $\bm{M}$ $(\times 10^{-3})$$\downarrow$ & \textbf{CV}$\downarrow$ \\
\hline
Crest-factor-optimised multisine & 2453          & 18.79          & 0.71          & 1.09          \\
Swept sine                       & 3101          & 44.97          & \textbf{0.67} & 1.44          \\
Space-filling multisine          & 2820          & 47.34          & 0.92          & 1.99          \\
RL input                         & \textbf{4820} & \textbf{71.72} & 1.84          & \textbf{0.52} \\
\hline
\end{tabular}
\end{table}

\subsubsection{Van der Pol system}
\begin{figure}[p]
  \centering
  \includegraphics[width=\textwidth]{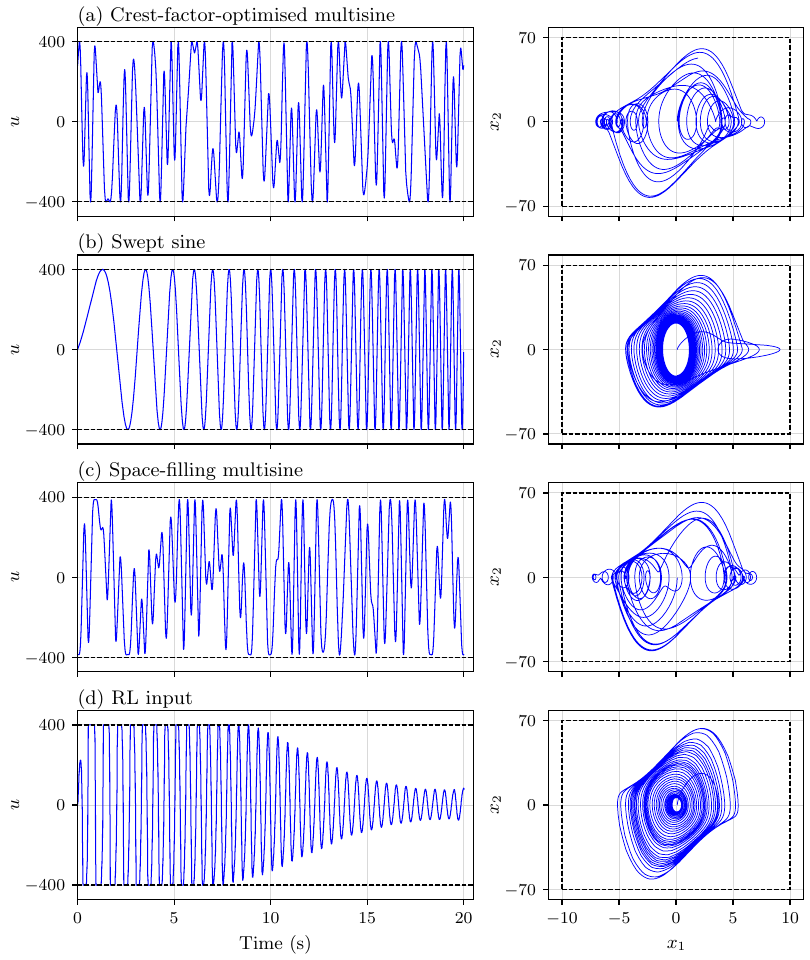}
  \caption{Input signals (left) and the state-space trajectories they induce (right) on the Van der Pol system. The dashed lines mark the amplitude bound and the region of interest.}
  \label{fig:vdp_signals}
\end{figure}

Fig.~\ref{fig:vdp_signals} shows the input signals and state-space trajectories for the Van der Pol system, with the same layout as Fig.~\ref{fig:mech_signals}. Table~\ref{tab:perf_vdp} reports the corresponding metrics.

Unlike on the mechanical system, optimising the phases for space filling brings only a limited improvement. The space-filling multisine covers about the same domain as the crest-factor-optimised multisine ($35.75\%$ against $36.93\%$), with the same coefficient of variation and a similar return. It mainly reduces the largest gap, with $M$ dropping from $2.91\times10^{-3}$ to $1.70\times10^{-3}$, consistent with a cost that penalises poorly covered regions.

The swept sine covers a domain of the same size as the RL input, $30.55\%$, with the second-lowest coefficient of variation but larger gaps ($M = 2.75\times10^{-3}$). Its frequency range has, however, been tuned to maximise space-filling performance, and a non-optimised range leads to significantly worse results.

The RL input covers a smaller domain than the two multisines, $30.46\%$, but achieves the highest return and the lowest $M$ and $\mathrm{CV}$ of the four inputs. Its $M$ is about half that of the crest-factor-optimised multisine and the swept sine, and $15\%$ lower than that of the space-filling multisine. On this system, adjusting only the phases of a fixed amplitude spectrum thus seems to offer little room to improve coverage, whereas the RL input trades a smaller domain for a much more even coverage.

As with the mechanical system, the agent autonomously learns physically meaningful excitation patterns (Fig.~\ref{fig:vdp_signals}d), progressively reducing the amplitude to uniformly cover the inner region of the state space. In contrast to the mechanical system, the agent performs a single long amplitude decay rather than multiple amplitude sweeps. This highlights a key strength of the RL framework: the agent autonomously learns the form of the excitation itself and adapts it to each system, whereas the baselines only tune the parameters of a signal class chosen in advance, which may not suit the system dynamics.

\begin{table}[htbp]
\centering
\caption{State-space coverage performance on the Van der Pol system, read as in Table~\ref{tab:perf_mech}.}
\label{tab:perf_vdp}
\small
\begin{tabular}{lcccc}
\hline
\textbf{Input} & \textbf{Return}$\uparrow$ & $\bm{C}$ \textbf{(\%)}$\uparrow$ & $\bm{M}$ $(\times 10^{-3})$$\downarrow$ & \textbf{CV}$\downarrow$ \\
\hline
Crest-factor-optimised multisine & 2751          & \textbf{36.93} & 2.91          & 1.77          \\
Swept sine                       & 2910          & 30.55          & 2.75          & 1.43          \\
Space-filling multisine          & 2725          & 35.75          & 1.70          & 1.75          \\
RL input                         & \textbf{2955} & 30.46          & \textbf{1.45} & \textbf{1.28} \\
\hline
\end{tabular}
\end{table}

\section{Conclusions}\label{sec:conclusion}
This paper proposed a reinforcement learning approach for space-filling input design of nonlinear systems. Unlike conventional optimisation-based methods, the proposed framework does not rely on predefined excitation signal classes nor on system-specific frequency tuning, and instead learns to drive the system to uniformly explore its state space under amplitude constraints.

Under the same amplitude bound, the learned policies achieve the best overall space-filling performance on both systems, improving on carefully tuned swept sine and crest-factor-optimised multisine excitations. The space-filling design of Kiss et al.~\cite{KISS2024} covers a much larger domain than the crest-factor-optimised multisine on the mechanical system, though not on the Van der Pol system, but still distributes its samples less uniformly than the RL inputs on both systems. The agent autonomously discovers physically meaningful excitation strategies adapted to the nonlinear dynamics of the system, such as resonance excitation and amplitude modulation. Beyond generating a single optimised excitation signal, the framework yields a policy that produces space-filling inputs for new initial conditions without re-optimisation, by conditioning the excitation on the current system observations rather than relying on a fixed open-loop signal.

These improvements have direct practical implications for nonlinear system identification. Achieving better state-space coverage within the same experiment duration is expected to yield more representative datasets and improved model generalisation. Additionally, for an equal level of coverage, the proposed approach has the potential to significantly reduce experiment time and associated costs, which is particularly relevant for engineering applications.

Future work will identify nonlinear models from the datasets generated by each input and compare their prediction errors on independent validation signals, in order to establish directly the link between state-space coverage and model quality. Training the agent on a surrogate model and applying the resulting input directly to the real system, in the presence of model mismatch and measurement noise, will also be investigated. A further step is to deploy the policy learned on the surrogate online on the real system, using a state encoder~\cite{Beintema2023} to estimate the states in its observation from the measured inputs and outputs.

%% ---------------------------------------------------------------

\section*{CRediT authorship contribution statement}
\textbf{Charly Witmeur:} Conceptualization, Methodology, Software, Formal analysis, Investigation, Visualization, Writing – original draft.

\textbf{Gonçalo G. Cruz:} Writing – review \& editing.

\textbf{Maarten Schoukens:} Writing – review \& editing, Supervision.

\textbf{Mark C. Runacres:} Supervision, Funding acquisition, Writing – review \& editing.

\textbf{Jan Decuyper:} Conceptualization, Supervision, Funding acquisition, Writing – review \& editing.

\section*{Declaration of competing interest}
The authors declare that they have no known competing financial interests or personal relationships that could have appeared to influence the work reported in this paper.

\section*{Declaration of generative AI and AI-assisted technologies in the manuscript preparation process}
During the preparation of this work, the authors used Claude (Anthropic) to assist in reformulating and improving the wording of text originally drafted by the authors. The authors reviewed and edited the resulting text as necessary and take full responsibility for the content of the published article.

\section*{Data availability}
Data will be made available on request.

\bibliographystyle{elsarticle-num}
\bibliography{ifacconf}

\end{document}